\documentclass{amsart}

\usepackage{eucal}

\usepackage{amssymb,amsxtra}

\usepackage{multicol}

\makeatletter

\def\theglossary{\@restonecoltrue\if@twocolumn\@restonecolfalse\fi
\columnseprule\z@ \columnsep 35\p@
\let\@makessectionhead\indexsec
\@xp\section\@xp*\@xp{\glossaryname}%
\let\item\@idxitem
\parindent\z@  \parskip\z@\@plus.3\p@\relax
\footnotesize}

\def\glossaryname{Notation Index}

\makeatother

\makeglossary

\numberwithin{equation}{section}
\renewcommand\dots{\relax\ifmmode\ldots\else$\,\ldots\,$\fi}

\newcommand\note[1]%
{$^\dagger$\marginpar{\sf\footnotesize{$^\dagger${#1}}}}

\divide\count253 by 60 
\divide\count255 by 60 
\multiply\count255 by 60 
\advance\count254 by -\count255 

\def\today{\number\day\space\ifcase\month\or January\or February\or
March\or April\or May\or June\or July\or August\or September\or
October\or November\or December\fi\space\number\year}

\def\hour{\ifnum\count253<10
0\number\count253\else\number\count253\fi}

\def\minute{\ifnum\count254<10
0\number\count254\else\number\count254\fi}

\swapnumbers

\newtheorem{proposition}[equation]{Proposition}
\newtheorem{lemma}[equation]{Lemma}

\theoremstyle{definition}

\newcommand\gt{\mathfrak{t}}
\renewcommand\gg{\mathfrak{g}}

\newcommand\gb{\mathfrak{b}}

\newcommand\bb[1]{{\text{\bf#1}}}

\newcommand\bbc{\bb{C}}

\newcommand\bbh{\bb{H}}
\newcommand\bbp{\bb{P}}
\newcommand\bbi{\bb{I}}

\newcommand\ca{\mathcal}

\newcommand\funclim[1]{\operatorname*{\mathrm{#1}}}

\renewcommand\lim{\funclim{lim}}

\begin{document} 
\title{On the Geometry of Isomonodromic Deformations}
\author{Jacques Hurtubise}
\address
{Department of   Mathematics and Statistics\\
McGill University \\
805 Sherbrooke St. W\\
Montreal H3A 2K6\\
Canada}
\email{Hurtubise@math.mcgill.ca}
\thanks{The author acknowledges the support of NSERC and FQRNT}


\begin{abstract}
This note examines the geometry behind the Hamiltonian structure of isomonodromy deformations of connections on vector bundles over Riemann surfaces. The main point is that one should think of an open set of the moduli of pairs $(V,\nabla)$ of vector bundles and connections as being obtained by ``twists" supported over points of a fixed vector bundle $V_0$ with a fixed connection $\nabla_0$; this gives two deformations, one,  isomonodromic, of $(V,\nabla)$, and another induced from the  isomonodromic  deformation of $(V_0,\nabla_0)$. The difference between the two will be Hamiltonian.
\end{abstract}
\maketitle

\section{Introduction}

The space of flat, or holomorphic, or more generally meromorphic connections on a Riemann surface has a rich geometric structure, tying in with many objects of interest in algebraic geometry. One need only think of   flat connections for a compact group $G$, and their link via the theorem of Narasimhan-Seshadri \cite{Narasimhan-Seshadri} with stable holomorphic $G_\bbc$-bundles; complexifying, there are links brought to light by Hitchin \cite{Hitchin} and Donaldson \cite{Donaldson} between flat $G_\bbc$-connections and the space of stable pairs. Allowing poles and restricting to the Riemann sphere, the theory has a long history going back to the beginning of the 20th century, with the Riemann-Hilbert problem of knowing whether a representation of the fundamental group of the sphere minus a certain number of punctures can be realised as the monodromy of a meromorphic connection with poles at the punctures; see, e.g., Bolibruch \cite{Bolibruch}. 

The fundamental group of a punctured Riemann surface is of course independent of its modulus, and so it is of interest to understand how the meromorphic connection associated to the representation varies as one deforms the modulus; i.e., one fixes the representation, and changes the modulus. The equations governing these isomonodromic deformations have been  studied since the early 20th century; special cases include the Schlesinger equations \cite{Schlesinger}, governing connections on the punctured Riemann sphere with simple poles, and the Painlev\'e equations \cite{Ince}, arising from connections for rank two bundles on the four-punctured sphere. The equations have many interesting features, including the Painlev\'e property (in essence, that, away from some subvarieties, the only singularities of solutions are poles). The latter has been best explained in work of Malgrange \cite{Malgrange}, who uses the full freedom of the description of bundles in terms of transition functions to solve the flows ``explicitly". 

When one allows connections with more than simple poles, the theory acquires additional complexity; the monodromy around the poles then decomposes into a number of Stokes factors; the eigenvalues of the higher order parts of the poles become invariants, to be thought of as generalised moduli, so that one can deform them also, as part of generalised isomonodromy deformations.

One of the main features of the moduli spaces of complex connections, developed in recent work of Hichin and Boalch \cite{Hitchin, Boalch} (see also Woodhouse\cite{Woodhouse}), is that they are endowed with complex symplectic forms, or in some cases, with complex Poisson structures. In parallel, over the Riemann sphere, (see, e.g. \cite{Harnad, Hitchin, Boalch}), the isomonodromy deformations have been given an interpretation as (non-autonomous) Hamiltonian flows.
In a remarkable paper, these Hamiltonian flows  were generalised by Krichever to arbitrary Riemann surfaces (for the ordinary moduli, for $Gl(n)$-connections, and for bundles of  degree $n(g-1)+n$). He uses for this the Tjurin parametrisation of vector bundles given by their global sections, pulling back the connections in effect to the fixed, trivial bundle. The constructions are very explicit, and indeed in some sense computationally effective. Another construction, due to Levin and Olshanetsky, studies the problem as an infinite dimensional reduction \cite{Levin-Olshanetsky}.

The Hamiltonians involved are, in all cases, invariants derived from the spectrum of the connection matrix. A priori, this is quite astounding: indeed, the notion is not at all gauge invariant, as one can gauge connections to zero. Another more minor puzzle is that one proceeds in general by analogy with the Hitchin systems: these have Hamiltonians which are invariant polynomials of arbitrary order in what corresponds to the connection matrix; here, for isomonodromic deformations correponding to variations in moduli  of the punctured curves one only considers the quadratic invariants. The aim of this paper is to elucidate the geometry behind these constructions, and generalise them to arbitrary degrees and to generalised monodromy. 

Our basic idea is quite simple: generalising the Tjurin parametrisation, one generates the generic bundle $V$ of a given degree from a fixed vector bundle $V_0$ by a series of ``twists" supported at points. In a similar way, from a fixed connection $\nabla_0$ on $V_0$, one can describe an arbitrary connection $\nabla$ on $V$ in terms of its polar parts (in an appropriate sense), supported at the same points. This point of view gives us two things: the first is that the difference $\nabla-\nabla_0$ has a well defined (form-valued) spectrum, and so provides us with our Hamiltonians, and the second is that the connection $\nabla_0$ has its own isomonodromic deformation, and carries along with it the pairs $(V,\nabla)$, through their polar parts. It is the difference between this flow and the isomonodromic flow which turns out to be Hamiltonian.

In section 2, we describe the spaces of bundles we consider, and in section 3, the spaces of connections. Section 4 modifies our constructions to deal with connections with poles. Section 5 builds the space in which the isomondromic deformations take place, and section 6 describes their Hamiltonian nature. In the last section, we make the flows autonomous, highlighting the role of quadratic differentials.

This paper follows on an earlier unpublished paper written together with Marco Bertola, John Harnad and Gabor Puzstai explaining the Hamiltonian nature of the generalised isomonodromy for higher order poles over the Riemann sphere. Their insight is gratefully acknowledged.

\section{ A configuration space; generalised Tjurin parameters}

Let $\Sigma$ be a compact Riemann surface of genus $g$, and let
$V_0$ be a fixed vector bundle over $\Sigma$ of degree $k_0$ and rank $n$. Consider the sheaf of groups $Aut(V_0)$, and the sheaf of modules
$Hom^0(V_0, V_0)$ consisting of generically invertible homomorphisms. One can build the quotient sheaf (of pointed sets) 
\begin{equation}{\ca F}= Hom^0(V_0, V_0)/Aut(V_0),
\end{equation}
 the action being given by $T\mapsto TS^{-1}$. Sections $T$ of this sheaf are supported at the points at which the determinant of $T$ vanishes.  The sections of ${\ca F}$ over $\Sigma$ have multiple components, given by the order of vanishing of the determinant, They map to the family ${\ca C}$ of pairs (bundles $V$, generically invertible $\phi:V \rightarrow V_0$). Indeed, for a section $T$ in $ H^0(\Sigma, {\ca F})$, this is done by covering $\Sigma$ by two open sets $U_0$ and $U_1$,  with $U_0$ defined as the complement of the support of $T$, and $U_1$ a disjoint union of disks around the finite set of points in the support of $T$. Now define a bundle $V$ as a subsheaf of $V_0$,   identical to $V_0$ on $U_0$, and given as the image of $T$ over $U_1$. Note that $T$ and $TS^{-1}$, for $S$ invertible,  have the same image. Alternately, one can use the section $T^{-1}$ as a transition matrix, as follows. The diagram 
\begin{equation}
\begin{matrix}
V&\buildrel{T^{-1}}\over{\longrightarrow}&V\\
I\downarrow&&\downarrow T\\
V_0&\buildrel{I}\over{\longrightarrow}&V_0
\end{matrix}
\end{equation}
defines the bundle $V$ and the map to $V_0$.
Here the left hand side is the isomorphism of $V$ with $V_0$ over $U_0$, and the right hand side is the map over $U_1$. The horizontal maps represent transition functions.  This is essentially the construction of bundles by ``matrix divisors", going back to Weil\cite{Weil}; see also Tjurin \cite{Tjurin}. One has an exact sequence
\begin{equation}
0\rightarrow V \rightarrow V_0 \rightarrow V/V_0 \rightarrow 0,\label{basic}\end{equation}
with the sheaf $V/V_0$ having the same support as $T$. We let the {\it degree} of $T$, $d= d(T)$, be the dimension $h^0(\Sigma, V/V_0)$ and denote by $H^0_d(\Sigma,{\ca F})$ the degree $d$ component of $H^0(\Sigma,{\ca F})$. For a section $T$ of ${\ca F}$, let $E=E(T)$ denote the divisor 
\begin{equation}E=\sum_{p\in \Sigma}
h^0(U_p, V/V_0)\cdot p,\end{equation}
 where $U_p$ is a small open set containing $p$ with at most $p$ belonging to the intersection of $U_p$ with the support of $T$. One has the easily proven lemma
\begin{lemma}
The degree of $T$ equals the difference of degrees $c_1(V_0)-c_1(V)$. It is also given by summing the orders of vanishing of the determinant of $T$ over its support, as well as by the degree of the divisor $E$.
\end{lemma}

The global automorphisms $H^0(\Sigma, Aut(V_0))$ act on ${\ca F}$ by $T\mapsto gT$, and it is clear that $T, gT$ define equivalent elements of ${\ca C}$. Thus:

\begin{proposition}
The quotient of the space of sections $H^0_d(\Sigma,{\ca F})$ by the action of $H^0(\Sigma, Aut(V_0))$ is isomorphic to ${\ca C}_{k_0-d}$, the subspace of ${\ca C}$ for which the bundle $V$ is of degree ${k_0-d}$.
\end{proposition}

As in \cite{Tjurin}, we can study a generic section of ${\ca F}$, showing that it consists of a sum of multiplicity one local sections located at distinct points. For points of multiplicity one, there is a simple invariant, which is the $n-1$-dimensional image of the matrix $T_i(z)$ at points $p_i$ where the determinant vanishes; this gives an element of the projectivisation $\bbp^{n-1}_{p_i}$ of the dual of the fiber of $V_0$ at $p_i$. This hyperplane, and the point $p_i$ where the determinant vanishes, determine the local section of ${\ca F}$, giving $n$ parameters in all for each point. There is a normal form for the matrix of the generic  degree one local section:
\begin{equation}
\begin{pmatrix}z-T_1&-T_2&-T_2&...&-T_n\\0&1&0&...&0\\0&0&1&...&0\\.&.&.&&.\\.&.&.&&.\\0&0&0&...&1
\end{pmatrix}\end{equation}
Here $z$ is some local coordinate on $\Sigma$.

Now fix the degree $k$ of $V$. We would like to describe some large family of bundles as the $V$s obtained from sections of ${\ca F}$ attached to a fixed $V_0$.  As the dimension of the space of stable bundles is $n^2(g-1) +1$, we would like a family of this dimension. Suppose  that $k +n(g-1) + 1= nt +s$, so that $k=(s-1)$(mod $n$). One can choose $V_0$ as a sum 
\begin{equation}
V_0= \oplus_{i=1}^n L_i,\ {\rm with}\  deg(L_i) = (t+1)(i\leq s)\ {\rm or}\ t (i>s)\ {\rm and}\ L_i\ {\rm distinct}.\label{basebundle}
\end{equation}

 Choosing $n(g-1) + 1$ points $p_i$ in the surface, and putting multiplicity one sections $T_i(x)$ of ${\ca F}$ at these points, gives an $(n^2(g-1) + n)$-parameter family of sections of ${\ca F}$; quotienting by the global automorphisms $\bbc^n$ of $V_0$ gives then an $n^2(g-1) + 1$-parameter family of bundles, as desired. These are of degree $deg(V_0) - n(g-1) - 1$.  

(If $k +n(g-1) + n = nt$, there is another approach: one can choose $V_0$ to be the trivial bundle tensored with a line bundle of degree $t$, and take $n(g-1) + n$ points. The automorphisms are now $n^2$-dimensional, and one again obtains an $(n^2(g-1) + 1)$-parameter family of bundles, now of degree $deg(V_0) - n(g-1) - n$. This is in essence the case treated by Tjurin. )

One would like to understand the family of bundles produced in this way. On the level of tangent spaces, to understand the deformations, one takes the sequence \ref{basic}, and tensors it with $V^*$:

\begin{equation}
0\rightarrow V^*\otimes V \rightarrow  V^*\otimes V_0 \rightarrow  V^*\otimes V_0/V \rightarrow 0,\label{bundlesequence}\end{equation}

The coboundary $H^0(\Sigma, V^*\otimes V_0/V)\rightarrow 
H^1(\Sigma, V^*\otimes V)$ gives  the deformations of bundles associated to a deformation of $T$, though of as a map $V\rightarrow V_0$.  On the other hand, one can also take the sequence 
\begin{equation}
0\rightarrow V_0^*\otimes V_0 \rightarrow  V^*\otimes V_0 \rightarrow  V^*/V_0^*\otimes V_0 \rightarrow 0,\end{equation}
which gives a map $H^0(\Sigma, V_0^*\otimes V_0) \rightarrow  H^0(\Sigma, V^*\otimes V_0)$; this is the infinitesimal version of the action of the automorphisms of $V_0$; it maps trivially to the deformations of the bundles, giving
\begin{equation}
T_{V}{\ca C}_k = H^0(\Sigma, V^*\otimes V_0/V)/ H^0(\Sigma, V_0^*\otimes V_0)\rightarrow 
H^1(\Sigma, V^*\otimes V)\label{deformation}
\end{equation}
as our tangent map.

More globally, we want to understand whether or not the generic bundle constructed in this way is stable. In \cite {Tjurin}, it was shown that the general subsheaf $V$ of ${\ca O}^{\oplus n}$ constructed as above is indeed stable; the proof proceeds in essence by showing that this process gives an $n^2(g-1)+1$ dimensional family of distinct bundles, and that unstable bundles all come in families of lower dimension, using the Hardy-Narasimhan filtration of the bundle. The same procedure works here in general. Given Tjurin's results, it suffices to show that the infinitesimal deformation map \ref{deformation} is onto, at a generic point.  

\begin{proposition} a) Let $g\geq 2$. Let ${\ca C}_k$ be one of the $n^2(g-1) + 1$-parameter families of bundles constructed above from a sum  \ref{basebundle} of generic, distinct line bundles.  For a generic choice of a section $T$ of ${\ca F}$, the map from the tangent space   of ${\ca C}_k$ to the space $H^1(\Sigma, V^*\otimes V)$ of infinitesimal deformations of $V$ is an isomorphism of $n^2(g-1) +1$ dimensional spaces;  as deformations on curves are unobstructed, this gives an $n^2(g-1)+1$-dimensional family, whose generic element must be stable.  

b) At these points, one has an isomorphism $H^0(\Sigma, V^*\otimes V_0)\simeq H^0(\Sigma, V_0^*\otimes V_0)$

c) Conversely, let  $V\in {\ca B}_k $, the space of stable bundles. Then, for $V$ generic,  $H^0(\Sigma,Hom(V, V_0))$ is at least of dimension one and its elements are generically (over $\Sigma$)  isomorphisms,  and so $V$ lies in the image of $H^0_d(\Sigma,{\ca F})$.

\end{proposition}

As noted, our tangent map was induced from the coboundary
\begin{equation}
H^0(\Sigma, V^*\otimes (V_0/V))\rightarrow H^1(\Sigma, V^*\otimes V)
\end{equation}
 We show that the map is onto at special points, and so onto generically. The points we choose all have $T$s of the form $diag(1,...,1,z,1,..1)$, where $z$ is a coordinate on the curve. This creates a bundle $V$ of the form $\oplus_{i=1}^n L_i(-D_i)$, where the $D_i$ are positive sums of distinct points, and have
disjoint support. We can arrange things so that the degrees of the $L_i(-D_i)$ all lie within one of each other.

Let $p_i$ be a point of $D_i$, so that $T(z)$ at $p_i$ has the form $diag(1,..,1,z,1,..,1)$, with the $z$ in the $i$th position. Deforming $T$ by $T+\epsilon t$, so that $t$ (holomorphic in $z$) represents a local section of $ V^*\otimes (V_0/V))$, we see that the corresponding cocycle $T^{-1}t$ in $V^*\otimes V = \oplus_{s,t=1}^n L_s^*L_t(D_s-D_t)$ is trivial for $s\neq i$, and has a simple pole in the natural trivialisations for $s=i$. The question then becomes one of whether these cocycles generate the first cohomology of the line bundles in a linearly independent fashion, which is true as one can choose the bundles $L_i$ generically.

One then has a surjection of our space of cocycles onto $H^1(\Sigma, V^*\otimes V) = \bbc^{n^2(g-1)+n}$; quotienting by the $n$ dimensional group of automorphisms of $V$, this gives the desired ${n^2(g-1)+1}$-dimensional family of bundles, which must include  stable bundles.

For b), one then has from the long exact sequence that $H^1(\Sigma, V^*\otimes V_0) = 0$; Riemann-Roch then tells us that $H^0(\Sigma, V^*\otimes V_0) $ is $n$-dimensional, and so isomorphic to $H^0(\Sigma, V_0^*\otimes V_0) $

For c), one simply computes $H^0(V^*\otimes L_i)$, using Riemann-Roch. The non-vanishing of the determinant can be seen from our special case of $\oplus_{i=1}^n L_i(-D_i)$, for a generic choice of $D_i$; the same must hold nearby.

\section{Phase spaces}

We now want to have phase spaces of twice the dimension of our configuration spaces ${\ca B}_k$ of stable bundles or ${\ca C}_k$ of bundles built as subsheaves of $V_0$. These spaces will be spaces of pairs of bundles and  connections.

We begin with ${\ca C}_k$. We first define a larger space than the one we want. Suppose we have, on an open set $U$,  an inclusion $V\rightarrow V_0$, corresponding to a section $T$ of ${\ca F}$. We will deem two connections 
$\nabla,\nabla'$ on $V$, defined over $U$, to be equivalent, ($\nabla\simeq\nabla'$), if their difference maps sections of $V_0$ into $V\otimes K_\Sigma$; note that holomorphic connections$\nabla,\nabla'$ on $V$ extend to  meromorphic connections on $V_0$, and we ask for equivalence not only that the difference be holomorphic, but that its image lie in $V$. Let ${\ca S}^V(U)$ denote the family of such equivalence classes; varying $U$ defines a sheaf ${\ca S}^V$  over $\Sigma$ ; global sections of ${\ca S}^V$ are supported over the support of $T$. Note that a global section of ${\ca S}^V$ need not correspond to an honest connection; it is just a configuration of equivalence classes localised at points. Set 
\begin{equation}
{\ca S}_k = \{((V\rightarrow V_0), (\nabla/\simeq))| (V\rightarrow V_0)\in {\ca C}_k, (\nabla/\simeq) \in H^0(\Sigma, {\ca S}^V)\}
\end{equation}

On an infinitesimal level, we note that the tangent space of $H^0(U, {\ca F})$ at $T$ is given by sections of $Hom(V, V_0)/Hom(V, V)= V^*\otimes (V_0/V)$, while the tangent space to the sections of ${\ca S}^V$ is given by sections of $Hom(V, V\otimes K_\Sigma)/Hom(V_0, V\otimes K_\Sigma) = (V^*/V_0^*)\otimes V\otimes K_\Sigma$. These two spaces of sections, both localised at the support of $T$, are dual: if $a$ represents a section of 
$Hom(V, V_0)/Hom(V,V)$ and $b$ a section of $Hom(V,V\otimes K_\Sigma)/Hom(V_0,V\otimes K_\Sigma)$ in a neighbourhood of the support of $T$, their pairing in the $U_1$ trivialisation is given by 
\begin{equation}
(a,b)\mapsto <a,b> = \sum res\ tr (bT^{-1}a)\label{pairing}
\end{equation}
(we are summing over the support of $T$) and in the $U_0$ trivialisation by
\begin{equation}
(a,b)\mapsto <a,b> = \sum  res\ tr (ba)\label{pairing0}
\end{equation}
We note that this is well defined; if we change the trivialisation of $V$ near the support of $T$ by $T
\mapsto TF$, then one has $a\mapsto aF, b\mapsto F^{-1}bF$, and the evaluation of the form does not change. It is also straightforward to see that it is non degenerate.

We now set our moduli space ${\ca R}_k\subset {\ca S}_k$, over ${\ca C}_k$, to be  a  space of  pairs on $(V\hookrightarrow V_0 , (\nabla/\simeq))$, with $\nabla$ a connection on $V$, i.e, the subspace corresponding to actual connections. As the degree of $V$ is not necessarily zero, we allow a single pole over a fixed point of $\Sigma$, with a fixed polar part; we will suppose that this base point is disjoint from the support of our sections of ${\ca F}$. We fix the residue of this connection around the fixed point to be $ (2k\pi i/n) \cdot \bbi$,  with corresponding monodromy $ exp (2k\pi i/n) \cdot \bbi$.

For $V$ generic, a connection $\nabla$ will be the only connection in its equivalence class: one has the exact sequence, dually to \ref{bundlesequence}

\begin{equation}
0\rightarrow V_0^*\otimes V\otimes K_\Sigma \rightarrow  V^*\otimes V\otimes K_\Sigma \rightarrow  V^*/V_0^*\otimes V\otimes K_\Sigma \rightarrow 0,\label{connectionsequence}\end{equation}
giving 

\begin{align}
0&\rightarrow H^0(\Sigma, V_0^*\otimes V\otimes K_\Sigma) \rightarrow  H^0(\Sigma, V^*\otimes V\otimes K_\Sigma) \rightarrow  H^0(\Sigma,V^*/V_0^*\otimes V\otimes K_\Sigma)\label{connectioncohomology}\\ & \rightarrow H^1(\Sigma, V_0^*\otimes V\otimes K_\Sigma)\notag \end{align}

The second term parametrises connections on the same bundle $V$, and the third the set of polar parts; the  connections with the same polar parts are parametrised by the first term, which is the dual of $ H^1(\Sigma, V^*\otimes V_0)$. The Riemann Roch theorem tells us that $ h^0(\Sigma, V^*\otimes V_0)-h^1(\Sigma, V^*\otimes V_0) $ is $n$, while  $ h^0(\Sigma, V^*\otimes V_0)$ is generically equal to $ h^0(\Sigma, V_0^*\otimes V_0) = n$, so the first term in \ref{connectioncohomology} is generically  zero, telling us that the map that associates a connection to its ``polar part" is injective. 

Globally, summing over the support of $T$, one has on the level of sections the duality
\begin{equation} H^0(\Sigma, V^*\otimes V_0/V) = H^0(\Sigma,(V^*/V_0^*) \otimes V\otimes K_\Sigma)^*.\end{equation}
Restricting to the image of $H^0(\Sigma, V^* \otimes V\otimes K_\Sigma)$ in $H^0(\Sigma,(V^*/V_0^*) \otimes V\otimes K_\Sigma)$ gives 
\begin{equation} H^0(\Sigma, V^*\otimes V_0/V)/ H^0(\Sigma, V^*\otimes V_0) = H^0(\Sigma,V^* \otimes V\otimes K_\Sigma)^*.\end{equation}
 (For this, one must recall that the sum of the residues of a globally defined meromorphic one-form is zero.) This identifies $H^0(\Sigma,V^* \otimes V\otimes K_\Sigma)$ as the cotangent space of ${\ca C}_k$.

  One  can use the pairing to define a two-form on ${\ca R}_k$, over the points of ${\ca C}_k$ for which $H^0(\Sigma, V^*\otimes V_0)\simeq H^0(\Sigma, V_0^*\otimes V_0)$. At these points, the tangent space to ${\ca C}_k$ is indeed given by $H^0(\Sigma, V^*\otimes V_0/V)/ H^0(\Sigma, V^*\otimes V_0)$, while the tangent space to the space of connections is $H^0(\Sigma, V^* \otimes V\otimes K_\Sigma)$. 

To define the pairing, we now suppose given a connection $\nabla_0$ on $V_0$; we suppose that it preserves the line bundles which are the summands of $V_0$. Again, if the degree of $V_0$ is non-zero, we allow a fixed simple pole with fixed residue at a fixed point disjoint from the divisors $E$ and $D$.

Taking covariant constant sections with respect to $\nabla_0$ defines a class of trivialisations of $V_0$ and so of $V$ over $U_0\cap U_1$, up to the action of constant matrices. Now if we have a connection $\nabla$ on $V$, there is in a similar fashion a natural class of trivialisations of $V$ over $U_1$. Relating the two trivialisations give natural choices of the matrices $T$ defining $V\rightarrow V_0$, defined up to the action of constant matrices. Thus a tangent vector to ${\ca R}_k$ gets represented by matrices $t$ representing the variation in $T$ and $b$ representing the variation in $\nabla$ in the $U_0$ trivialisation defined near the support of $T$ up to the action of constant matrices; over the punctured disks of $U_0\cap U_1$ near $E$, one has $\nabla (T^{-1}t) = b$.
Then, if $(t_1,b_1), (t_2,b_2)$ represent two such tangent vectors to ${\ca R}_k$, the symplectic form is defined as

\begin{equation}
  \omega((t_1,b_1), (t_2,b_2)) = <t_1,b_2>- <t_2,b_1>\end{equation}

\begin{proposition}
The variety ${\ca R}_k$ is symplectic, over the smooth locus of ${\ca C}_k$
\end{proposition}

The form is well defined, and non-degenerate; what remains to be seen is that it is closed. Let $d$ denote the exterior derivative along the curve, and $\delta$ the exterior derivative on ${\ca R}_k$. The form, as defined on ${\ca R}_k$, is locally the pull back of a form defined over the space of holomorphic matrix valued functions on the curve (modulo the actions of constant matrices, as above): indeed, one associates to a connection $\nabla$ its matrix $A$ in the $\nabla_0$ trivialisation; the corresponding transition function $T$ is given by solving $A = -\frac{dT}{dz} T^{-1}$. The form is then as in \cite{Krichever}
\begin{equation}res\ tr (-\delta (dT T^{-1})\wedge \delta T T^{-1}).\end{equation}
  Taking exterior derivative $\delta$, one finds
\begin{equation} res\ tr (-d(\delta T T^{-1})\wedge \delta T T^{-1} \wedge \delta T T^{-1}) 
\end{equation}
 which is zero.

Now let us consider the analogous phase space over ${\ca B}_k$,   the moduli space of semi-stable rank $n$ bundles of degree $k$: let ${\ca P_k}$ be the space of pairs $(V,\nabla)$, where $V\in {\ca B}_k$, and $\nabla$ is a holomorphic connection on $V$. Again, if $k\neq 0$, we have to allow a fixed singularity in $\nabla$.

\begin{proposition}
The variety ${\ca P}_k$ is symplectic, over the smooth locus of ${\ca B}_k$.
\end{proposition}

This is the Atiyah-Bott symplectic structure, extended to complex gauge groups as in Boalch \cite{Boalch}. One way of understanding this form in holomorphic terms is as follows: one has that the  tangent space to ${\ca P}_k$ is given by the first hypercohomology of the two step complex
$$Hom (V,V) \buildrel {\nabla}\over {\longrightarrow} Hom(V,V\otimes K). $$
(A cocycle with respect to a covering $U_i$ for this first hypercohomology is given by a Cech 1-cocycle  $\mu_{ij}$ for $Hom (V,V)$, and a  Cech 0-cochain $\nu_i$ for $Hom(V,V\otimes K)$ with  $\nabla \mu_{ij} = \nu_i-\nu_j$ on $U_i\cap U_j$.) Dualising, the cotangent complex is the same, and so there is a natural symplectic structure induced by the identity map, as in Markman \cite{Markman}. This space of holomorphic connections can be thought of as a deformation of the Hitchin phase space of Higgs pairs \cite{Hitchin-stable}.

Just as we saw that there was a generic isomorphism between ${\ca B}_k$ and ${\ca C}_k$ when the degrees were correct, one has a generic isomorphism between ${\ca P}_k$ and ${\ca R}_k$. At the generic stable bundle, one has that the differential of the map from ${\ca C}_k$  to ${\ca B}_k$ is given by 
\begin{align} H^0(\Sigma, V^*\otimes V_0/V)/ H^0(\Sigma, V_0^*\otimes V_0) &= H^0(\Sigma, V^*\otimes V_0/V)/ H^0(\Sigma, V^*\otimes V_0)\\ &\rightarrow 
H^1(\Sigma, V^*\otimes V),\notag \end{align} which is generically an isomorphism, while for the connections, the derivative is the identity map. 

\begin{proposition}
This identification is symplectic.
\end{proposition}

Indeed, the explicit calculation of the symplectic form from the hypercohomology proceeds as follows. Covering the curve again by our two open sets $U_0,U_1$, we have that the connection matrices $A_0, A_1$ in the two trivialisations are related by $A_0 =-dT T^{-1} + TA_1 T^{-1}$. The form on ${\ca P}_k $ is then
$\sum res\ tr ((\delta A_0+\delta A_1)\wedge \delta T T^{-1})$, where one sums the residues over the support of $T$. The gauge chosen to describe the form on ${\ca R}_k$ is one on which $A_1$ vanishes.
Going to this gauge, one has that the formulae for the connections on ${\ca P}_k$ and ${\ca R}_k$ coincide. 

\section{  Poles. }

We   want to consider not only holomorphic connections, but also meromorphic connections with $\ell$ poles on a positive divisor 
\begin{equation}
D = \sum_{i=1}^\ell l_ip_i.
\end{equation}
We let $D_{red}$ be the  reduced divisor of $D$; i.e., if $D= \sum l_ip_i, l_1\geq l_2 \geq...\geq l_\ell, p_i\in \Sigma$, $p_i$ distinct, then $D_{red}= \sum p_i$. Set $L(D) = (l_1,l_2,...,l_\ell)$. We will suppose that the support $D$ is disjoint from the support $E$ of $T$. For the connections to exist, we no longer need the bundles to be of degree zero; correspondingly, if $D$ is non zero, we do not need to allow a fixed pole elsewhere. Spaces of bundles and meromorphic connections are no longer symplectic, but only Poisson;  in the by now standard way, following \cite{Markman, Jeffrey, Boalch}, we add some trivialisations over $D$ to our data, setting ${\ca B}_{k,D}$ to be the moduli space of rank $n$ bundles of degree $k$ equipped with a trivialisation over $D$. We similarly enlarge the space of pairs (bundles, connections) to 
\begin{equation} 
{\ca P}_{k,D}= \{(V,tr, \nabla)|  (V, tr) \in {\ca B}_{k,D}, \nabla\ {\rm a\ connection\ with\ polar\ divisor\ }\leq D\}.
\end{equation}
 As above,
\begin{proposition}
The variety ${\ca P}_{k,D}$ is symplectic, over the smooth locus of ${\ca B}_{k,D}$.
\end{proposition}

The tangent space to ${\ca P}_{k,D}$ at $(V,tr,\nabla)$ is now given by the first hypercohomology of the two step complex
\begin{equation} Hom (V,V)(-D)\buildrel {\nabla}\over {\longrightarrow} Hom(V,V\otimes K)(D),\end{equation}
and again the same complex yields the cotangent space, identifying the two, and defining the symplectic form. 

There is a symplectic action of $Gl(n)_D$, the group of maps from the scheme corresponding to $D$ into $Gl(n,\bbc)$;  the action is the natural one on the trivialisations over $D$; its moment map, into $(gl(n)_D)^*$, is given by the expression of the polar part of the connection in the chosen trivialisation. The reduced spaces ${\ca P}^\gamma_{k,D, red}$ are then symplectic spaces of pairs $(V, \nabla)$ with the polar parts of $\nabla$ lying in fixed conjugacy classes $\gamma$ over $D$.

For the Poisson manifold ${\ca P}_{k,D,red}= {\ca P}_{k,D}/Gl(n)_D = \cup_\gamma {\ca P}^\gamma_{k,D,red}$, one has the tangent complex, whose first hypercohomology is the tangent space: 
$$Hom (V,V)\buildrel {\nabla}\over {\longrightarrow} Hom(V,V\otimes K)(D),$$
and dually, the cotangent complex:
$$Hom (V,V)(-D)\buildrel {\nabla}\over {\longrightarrow} Hom(V,V\otimes K).$$
The natural injection of the latter complex into the first gives the Poisson structure.  

There is a ``partial" version of this reduction, considered by Boalch\cite{Boalch}. Indeed, one has within $Gl(n)_D$ the subgroup $B_D$ of maps whose leading term is the identity; this group is non-trivial only over the points of multiplicity at least two. Boalch shows how the corresponding moment map for this action is given by the ``irregular" polar part of the connection, i.e., the terms of order at most $-2$ in the Laurent expansion of the connection matrix. Reducing by the action of the group $B_D$, one thus is led to the ``partially" reduced spaces ${\ca P}^\gamma_{k,D, pred}$ of connections with fixed irregular type $\gamma$ (at least over the set where the leading order terms of these types is a regular element of the Lie algebra of the torus). In this space, one still has a trivialisation over $D_{red}$ and the order $-1$ term of the connection is free.

For ${\ca P}_{k,D,pred}= {\ca P}_{k,D}/B_D = \cup_\gamma {\ca P}^\gamma_{k,D,pred}$, one has the tangent complex, whose first hypercohomology is the tangent space: 
$$Hom (V,V)(-D_{red})\buildrel {\nabla}\over {\longrightarrow} Hom(V,V\otimes K)(D),$$
and dually, the cotangent complex:
$$Hom (V,V)(-D)\buildrel {\nabla}\over {\longrightarrow} Hom(V,V\otimes K)(D_{red}).$$
The natural injection of the latter complex into the first gives the Poisson structure.  

In another vein, one can restrict to trivialisations in which the polar parts are diagonal, and still have, over the regular elements,  a symplectic space ${\ca P}_{k,D,diag}$ (symplectic cross section). There is now an action of the group $T_D$ of maps from $D$ into the torus $T$, and so of the subgroup $T_D\cap B_D$; the moment maps for the action of $T_D$ is now into the dual space of $\gt_D$ consisting of ``polar parts" $t^i_{-l_i}z^{-l_i}+...+t^i_{-1}z^{-1}$ of maps into the Lie algebra of the torus; that for $T_D\cap B_D$ is into the the space of ``irregular polar parts" $t^i_{-l_i}z^{-l_i}+...+t^i_{-2}z^{-2}$. Reducing now by the actions of $T_D$, $T_D\cap B_D$, one gets the spaces ${\ca P}^\gamma_{k,D,red}$ again in the first case,  and  spaces ${\ca P}^{\gamma'}_{k,D,diag, pred}$ in the second; for the latter, the conjugacy class ${\gamma'}$ of the irregular polar part is fixed, with leading order term lying  in the algebra of the torus in the given trivialisations , while the order $-1$ terms are free.

We can, as in the previous section, consider the phase space corresponding to ${\ca C}_{k,D}$, where the bundles are  built from $V_0$, rather than ${\ca B}_{k,D}$, the moduli of stable bundles plus trivialisations. Thus, we set:
\begin{equation} 
{\ca R}_{k,D}= \{(V,tr, \nabla)|  (V, tr) \in {\ca C}_{k,D}, \nabla\ {\rm a\ connection\ with\ polar\ divisor\   }\leq D\}.
\end{equation}

This is again symplectic, and identifies symplectically with ${\ca P}_{k,D}$ over a large open set. To define the symplectic form, we note that in passing from ${\ca C}_{k}$ to ${\ca C}_{k,D}$, the tangent space gets enlarged from $H^0(\Sigma, V^*\otimes V_0/V)/ H^0(\Sigma, V^*\otimes V_0)$, supported at $E$, to $H^0(\Sigma, V^*\otimes V_0/V)/ H^0(\Sigma, V^*\otimes V_0) \times H^0(D,V_0^*\otimes V_0)$ supported at $E+D$. Meanwhile, the tangent space to the space of connections is $H^0(\Sigma, V^* \otimes V\otimes K_\Sigma(D))$. 

Choosing as above the  matrices $T$ defining $V\rightarrow V_0$ defining $V$ near $E$, let $t_i$ denote infinitesimal variations of these in $H^0(\Sigma, V^*\otimes V_0/V)/ H^0(\Sigma, V^*\otimes V_0)$; let $s_i\in H^0(D,V_0^*\otimes V_0)$ denote infinitesimal variations of the trivialisation, and let $b_i$ denote infinitesimal variations of the connections. Then the symplectic form applied to  $(t_1,s_1, b_1), (t_2,s_2, b_2)$ is given by 

\begin{equation}
  \omega((t_1,s_1, b_1), (t_2,s_2, b_2)) = <t_1,b_2>- <t_2,b_1> + res_Dtr(s_1b_2-s_2b_1)\end{equation}

The  residue at $D$ term above is essentially the canonical form on the cotangent bundle of the groups of maps from $D$ into $Gl(n,\bbc)$, and so, referring to our previous result: 

\begin{proposition}
The variety ${\ca R}_{k,D}$ is symplectic, over the smooth locus of ${\ca C}_{k,D}$.
\end{proposition}

As before, 

\begin{proposition}
The natural identification of ${\ca P}_{k,D}$ and ${\ca R}_{k,D}$ over a large open set is symplectic.
\end{proposition}

Finally, taking the difference $A= \nabla-\nabla_0$ with $\nabla_0$ thought of as a connection on $V$, we can think of  connections on $V$  as sections of $End(V)\otimes K_\Sigma$ with poles at $D+E$. For pairs  $(V,\nabla)$ thought of in this way, one has a deformation complex that is a subcomplex of
\begin{equation}Hom (V,V)(-D)\buildrel {[\cdot,A]}\over {\longrightarrow} Hom(V,V\otimes K)(D + 2E),\end{equation}
in that the order two term of the pole at $E$ is forced to be that of $T^{-1} (\delta T)T^{-1}dT$, where $T$ is the transition matrix considered above.
 
\section{  Enlarging to moduli.}

We would now like to vary the base curve $\Sigma$, or the location of the poles, thought of as punctures on $\Sigma$, or, when the poles are irregular, the conjugacy class of the higher order polar parts; these are, after all, the parameters for isomonodromic deformations.  We restrict to the set of connections with regular principal part, i.e. those connections whose  poles have a regular leading order term, conjugate to an element of the maximal torus. 

More precisely, let $L$ be a decreasing sequence $\{l_1\geq l_2\geq ...l_\ell>0 \}$ of positive integers. We will also occasionnally use $L$ to denote the associated standard scheme consisting of  the disjoint union over $i$ of the $(l_i-1)$-th formal neighbourhoods of the origin in $\bbc$, equipped with the standard coordinate. Now set
\begin{equation}
{\ca U}_{g, L, k} = \{(\Sigma, D, c, V, tr, \nabla)\}
\end{equation}
where $\Sigma$ is a Riemann surface of genus $g$, $D$ is a divisor on $\Sigma$ with $D_{red}$ of degree $\ell$ and $L(D) = L$, $c$ a set of  jets of coordinates on $\Sigma$ centered at the points of $D$, with an $l_i-1$-jet at the point  of multiplicity $l_i>1$, $V$ a stable vector bundle over $\Sigma$ of rank $n$ and degree $k$, $\nabla$ a connection on $V$ with polar divisor $D$ and regular principal part, $tr$ a trivialisation of $V$ over $D$. One has the subspace 
\begin{equation}
{\ca U}_{g, L, k, diag} = \{(\Sigma, D, c, V, tr, \nabla)\}
\end{equation}
where one restricts to trivialisations in which the polar parts of the connection are diagonal. Note that the jet $c$ allows us to think of $(\Sigma, D, c)$ as a pair consisting of $\Sigma$ and an inclusion morphism $L\mapsto \Sigma$

Now let ${\ca M}_{g,L}$ be the moduli of Riemann surfaces $\Sigma$ of genus $g$ with $\ell$ punctures at the support of a divisor $D$; at the i-th point, one has in addition a centred coordinate to order $l_i-1$, if $l_i>1$. Over ${\ca M}_{g,L}$ one can build the bundles of groups  ${\ca M}(GL(n)_D)$, ${\ca M}(B_D)$, ${\ca M}(T_D)$ or ${\ca M}(B_D\cap{T}_D)$, and the corresponding bundles of Lie algebras ${\ca M}({\gg}(n)_D)$, ${\ca M}({\gb}_D)$, ${\ca M}({\gt}_D)$ or ${\ca M}({\gb}_D\cap{\gt}_D)$, and the dual bundles ${\ca M}({\gg}(n)_D^*)$, ${\ca M}({\gb}_D^*)$, ${\ca M}({\gt}_D^*)$ or ${\ca M}(({\gb}_D\cap{\gt}_D)^*)$, all over ${\ca M}_{g,L}$; just as elements of the Lie algebra bundle can be thought of as truncated Taylor series (maps from the scheme corresponding to $D$ into the Lie algebra), elements of these dual bundles can be thought of as polar parts, or truncated Laurent series, of Lie algebra valued 1-forms; thus at a point of $D$ of multiplicity $\ell$, an element of ${\ca M}({\gt}_D^*)$ gets represented by a series $\sum_{i=-\ell}^{-1}t_iz^i$; elements of ${\ca M}(({\gb}_D\cap{\gt}_D)^*)$ by a series  $\sum_{i=-\ell}^{-2}t_iz^i$. Finally, for ${\ca M}(({\gb}_D\cap{\gt}_D)^*)$, one can quotient by the  group $W_D$ (maps from $D-D_{red}$ into the Weyl group $W$), to obtain the bundle ${\ca M}(({\gb}_D\cap{\gt}_D)^*/W_D)$ of conjugacy classes of irregular polar parts.  Note that the jet of coordinate $c$ trivialises   these bundles over ${\ca M}_{g,L}$, by giving an isomorphism of $GL(n)_D$, etc.  with the standard schemes $Gl(n)_L$, etc., so that for example:
\begin{align}
{\ca M}(GL(n)_D)&= {\ca M}_{g,L}\times GL(n)_L,\label{bundle-trivial}\\
{\ca M}(({\gb}_D\cap{\gt}_D)^*/W_D)&= {\ca M}_{g,L}\times ({\gb}_L\cap{\gt}_L)^*/W_L\notag \cr 
\end{align}
with the same holding for the other bundles.

For a connection $\nabla$, let $irr(\nabla)$ represent the conjugacy class of its irregular polar parts, and $polar(\nabla)$ the conjugacy class of the full polar part.
One has then a fibering 
\begin{align}
{\ca U}_{g, L, k } &\rightarrow  {\ca M}_{g,L}\times ({\gb}_L\cap{\gt}_L)^*/W_L\label{fibering}\\
(\Sigma, D, c, V, tr, \nabla)& \mapsto ((\Sigma, D, c), irr(\nabla))\notag 
\end{align}
which is invariant under  the action of the group
$B_L$; quotienting by the action  gives
\begin{equation}
\Pi: {\ca U}_{g, L, k }/B_L\  \rightarrow  {\ca M}_{g,L}\times ({\gb}_L\cap{\gt}_L)^*/W_L\label{fibering2}
\end{equation}
The fibers of \ref{fibering2} are the partial reductions ${\ca P}^\gamma_{k,D, pred}$, and so are symplectic. 

 The isomonodromy deformations give splitting of these maps, i.e. local lifts ${\ca M}_{g,L}\times ({\gb}_L\cap{\gt}_L)^*/W_L\rightarrow {\ca U}_{g, L, k}/B_L $. In other words, they give a connection on the space, and one has, by the results of  Hitchin, Boalch \cite{Hitchin, Boalch}:
\begin{proposition}
The isomonodromy flows preserve the symplectic form on the fibers, i.e. the connection is symplectic.
\end{proposition}
The geometric origins of these isomonodromic splittings is quite clear. Indeed, note that  a connection $\nabla$ gives  a local system on $V$. Deformations of the moduli of the punctured curve correspond to cutting the curve into patches, and reglueing these differently. These operations, for one-parameter deformations starting at the identity,  lift naturally  to the local systems on $V$. The isomonodromy deformation as one varies the irregular principal parts can be understood in the same way, in terms of glueing of local systems.

Similarly, one has
\begin{align}
{\ca U}_{g, L, k }/ Gl(n)_L &\rightarrow  {\ca M}_{g,L}\times ({\gt}_L)^*/W_L\label{fibering3}\\
(\Sigma, D, c, V, tr, \nabla)& \mapsto (\Sigma, D, c), polar(\nabla))\notag 
\end{align}with fibers the full reductions ${\ca P}^\gamma_{k,D, red}$.  The connection descends to this space.

On the level of tangent bundles, the symplectic connection can be understood as follows. We restrict to a simple case, to illustrate. Consider the tangent bundle of the principal bundle associated to $V$. This is a $Gl(n)$-equivariant bundle. Over $\Sigma$, its quotient is the Atiyah bundle \cite{Atiyah}, an extension
\begin{equation}
0\rightarrow Hom (V, V) \rightarrow At \rightarrow T\Sigma \rightarrow 0 \label{atiyahbundle}
\end{equation}
 One has
the natural deformation sequence:
\begin{equation}
 H^1(\Sigma, Hom (V, V))\rightarrow H^1(\Sigma, At) \rightarrow H^1(\Sigma, T\Sigma) \rightarrow 0,\label{deformations}
\end{equation}
showing that $ H^1(\Sigma, At)$ indeed incorporates both deformations $H^1(\Sigma, Hom (V, V))$ of the bundle  and deformations of the curve $H^1(\Sigma, T\Sigma)$. Including both punctures and trivialisations over $D_{red}$ involves twisting the sequence above by ${\ca O}(-D_{red})$, giving the deformation sequence:
\begin{equation}
 H^1(\Sigma, Hom (V, V)(-D_{red}))\rightarrow H^1(\Sigma, At(-D_{red})) \rightarrow H^1(\Sigma, T\Sigma(-D_{red})) \rightarrow 0.\label{deformations2}
\end{equation}
The connection on $V$ defines a splitting of the sequence \ref{atiyahbundle}, inducing a corresponding splitting of the deformation spaces \ref{deformations}, \ref{deformations2}.

\section{Hamiltonian flows}

We have examined in previous sections the family of bundles ${\ca C}_k$ one could develop from a fixed bundle $V_0$. We assume again that the bundle $V_0$ has a fixed connection, possibly with a fixed singularity. Just as for our family of bundles $V$ with connection, the pair $(V_0, \nabla_0)$ has an isomonodromic extension as one deforms the curves, and so one can, at least locally, speak of $(V_0, \nabla_0)$ as being defined  over ${\ca M}_{g,L}$.

We now can replace, over $ {\ca M}( ({\gb}_D\cap{\gt}_D)^*/W_D)= {\ca M}_{g,L}\times ({\gb}_L\cap{\gt}_L)^*/W_L$,  our family of stable bundles ${\ca B}_k$ by the family of bundles ${\ca C}_k$ that we have created as subsheaves from the fixed bundle $V_0$. Correspondingly, our families of bundles plus connections plus trivialisations  ${\ca P}_{k,D},   {\ca P}_{k,D, diag}$ etc. become ${\ca R}_{k,D},   {\ca R}_{k,D, diag}$, etc.; instead of  spaces ${\ca U}_{g, L, k}$, ${\ca U}_{g, L, k, diag}$,etc.  we will have spaces ${\ca V}_{g, L, k}$, ${\ca V}_{g, L, k, diag}$,etc. The only thing that changes is the family of bundles.

At first glance, this doesn't change much, at least locally; the spaces of bundles coincide over a large open set. The difference is that we now have a background bundle $V_0$, and a background connection $\nabla_0$.  We can transfer all our connections $\nabla$ to the bundle $V_0$, and think of them as a connections there with extra singularities  at the support $E$ of the section $T$ of ${\ca F}$ defining $V$. (We suppose, as usual, that the supports of $E$ and $D$ are disjoint and that $E_{red} = E$.) The same transfer holds for the trivialisations, which can be transported to $V_0$. 

The background connection $\nabla_0$ does a very useful thing for us. Indeed, one now has, in addition to the  isomonodromic splittings $I: {\ca M}_{g,L}\times ({\gb}_L\cap{\gt}_L)^*/W_L
 \rightarrow {\ca V}_{g, L, k } /{\ca B}_L$  of $\Pi$,  defined by the connection $\nabla$ and isomonodromic for $\nabla$, a  second set of splittings $I_0$, which extend the isomonodromic splitting for $\nabla_0$ to other connections: one should think of the connection $\nabla$ and bundle $V$ being determined by their ``polar data" at the divisor $E+D$, and being carried along unchanged, supported, if one will, by the local system defined by $\nabla_0$ over $V_0$. The difference of the  two splittings $I$ and $I_0$, or rather of the vector fields defined from the corresponding connections, is a vertical vector field along the fibers of $\Pi$,  and will be given a Hamiltonian interpretation. 

 Cover, as before, $\Sigma$ by open sets $U_0$ consisting of $\Sigma$ minus the support of $D$ and $E$, and $U_1$ consisting of disks around the points of $D$ and $E$; we will subdivide $U_0$ and $U_1$ as needed into $U_{0,j}$ and $U_{1,j}$ respectively, for example when considering Stokes sectors near the points of $D$, or when choosing covariant constant bases with respect to $\nabla$ or $\nabla_0$. We use greek letters $\alpha, \beta$ to denote any of the indices $({0,j}),({1,i})$.

\begin{proposition}

a) Let $\mu\in H^1(\Sigma, T_\Sigma(-D))$ be a tangent vector to ${\ca M}_{g,L}$.  Let $v^{\mu}$, $v^{\mu}_0$ be their subsequent lifts to  ${\ca V}_{g, L, k}/{\ca B}_L $ under $I$, $I_0$ respectively.
Let $X_{\mu\cdot Q}$ be the Hamiltonian vector field on ${\ca R}^\gamma_{k, D, pred}$ corresponding to the function $H_{\mu\cdot Q}$, defined below. Then 

$$v^{\mu} - v^{\mu}_0 = X_{\mu\cdot Q}$$

b) Let $\beta\in({\gt}_L)^*$ be such that $d\beta\in ({\gb}_L\cap{\gt}_L)^*$ represents a tangent vector to $({\gb}_L\cap{\gt}_L)^*/W_L$. Let 
$v^\beta, v^\beta_0$ be the lifts of $d\beta$ to  ${\ca V}_{g, L, k }/{\ca B}_L $ under $I$, $I_0$ respectively. 
Let $X_{\beta\cdot B}$ be the Hamiltonian vector field on ${\ca R}^\gamma_{k, D, pred}$ corresponding to the function $H_{\beta\cdot B}$, defined below. Then 

$$v^\beta - v^\beta_0 = X_{\beta\cdot B}$$
\end{proposition}

We begin with lifts from ${\ca M}_{g,L}$. A tangent vector to ${\ca M}_{g,L}$ is given by an element of $H^1(\Sigma, T_\Sigma(-D))$; let $\mu_{10}(x)$ be a representative cocycle for a tangent vector with respect to the covering by $U_0$, $U_1$. We will suppose (simply to simplify notation) that it is concentrated on the disks around $D$, and vanishes on the disks around $E$. We then have $\mu_{(1,j)(0,i)}(x) = \mu_{10}(x)$, $\mu_{(0,j)(0,i)}(x)= \mu_{(1,j)(1,i)}(x) = 0$. The infinitesimal deformation in the Riemann surface consists in identifying  the point $x$ in $U_\alpha$ with $x+\epsilon\mu_{\alpha\beta}(x) $ in $U_\beta$.

We note first that we have projection maps
\begin{equation}H^1(\Sigma, T_\Sigma(-D-nE)) \rightarrow H^1(\Sigma, T_\Sigma(-D)),  \end{equation}
and that these can be split, as follows:
one has, associated to each connection $\nabla$, a transition matrix $T$ near $E$ from a $\nabla_0$-constant basis to a $\nabla$-constant basis, unique up to constant matrices. The function $\det (T)$ vanishes at $E$, and provides up to scale a uniquely defined coordinate $z$ near each point of $E$. This means that there is a well defined grading of order of vanishing in meromorphic sections of the tangent bundle near $E$, and so well defined ways of lifting cocycles for $T_\Sigma(-D)$ to $T_\Sigma(-D-nE)$ by asking that the terms of order $\frac{\partial}{\partial z}, ...., z^{n-1}\frac{\partial}{\partial z}$ in the lift vanish.

We now can define the Hamiltonians $H_{\mu\cdot Q}$, on each fiber ${\ca R}^\gamma_{k,D, pred}$. 
We have, for each point of the fiber ${\ca R}^\gamma_{k,D, pred}$ over $(\Sigma, D,c )$ in ${\ca M}_{g,L}$, a meromorphic  $End(V_0,V_0)$-valued form $(\nabla-\nabla_0)$, with poles at $D+E$ whose conjugacy class is fixed; there is then a meromorphic quadratic form $tr((\nabla-\nabla_0)^2)$, with polar divisor $2D +2E$, lying in $H^0(\Sigma, K_\Sigma^2(2D+2E))$. On each fiber ${\ca R}^\gamma_{k,D, pred}$ of 
\begin{equation}
{\ca V}_{g, L, k }/B_L \rightarrow  {\ca M}_{g,L}\times ({\gb}_L\cap{\gt}_L)^*/W_L, 
\end{equation}
 the image of $tr((\nabla-\nabla_0)^2)$ in $K_\Sigma^2(2D+2E))|_{D}$  is fixed. Choosing on each fiber a fixed $Q_0$ in $H^0(\Sigma, K_\Sigma^2(2D))$ with the same polar part, one has a well defined class $ Q = tr((\nabla-\nabla_0)^2)-Q_0$ belonging to $H^0(\Sigma, K_\Sigma^2(D+2E))$. 

Now let us take the lift of $\mu$ to $H^1(\Sigma, T_\Sigma(-D-2E))$ defined above, and set $H_{\mu\cdot Q}$ to be the pairing of $\mu$ with $Q$ in $H^1(\Sigma, K_\Sigma)$. Explicitly, 
\begin{equation}H_{\mu\cdot Q}= \sum_D res (\mu \cdot Q)\end{equation}
 Let $X_{\mu\cdot Q}$ be the corresponding Hamiltonian vector field on  ${\ca R}^\gamma_{k,D, pred}$.

The isomonodromic connection $I$ is simple to express, in trivialisations $f_{\alpha}$   that are covariant constant with respect to $\nabla$: one has constant patching functions $G_{\alpha,\beta}$ on the overlaps $V_\alpha\cap V_\beta$, and one keeps these patching functions constant as one varies the glueing of $V_\alpha$ and $V_\beta$ as one is moving along in  ${\ca M}_{g,L}$. The glueing of $\bbc^n\times V_\alpha$ to  $\bbc^n\times V_\beta$ is thus, to first order ($\epsilon^2 = 0$) :
 $$ (f_\alpha, x)\leftrightarrow  (G_{\alpha,\beta}f_\beta,x+\epsilon \mu_{\alpha,\beta}),$$
 Moving now to $\nabla_0$-constant trivialisations $e_\alpha $, related by constant matrices $H_{\alpha,\beta}$, we find that the identification is now given by 
$$ (e_\alpha, x)\leftrightarrow (H_{\alpha,\beta} (1+\epsilon\mu_{\alpha,\beta}A_{\beta}) e_\beta,x+\epsilon \mu_{\alpha,\beta})$$
where $A_\beta$ is the connection matrix for $\nabla$ (and so for $\nabla-\nabla_0$) in the $e_\beta$ trivialisation. The connection matrices $A_\beta$ are left constant over each $V_\beta$, with the change of gauge $(1+\epsilon\mu_{\alpha,\beta}A_\beta)$ compensating for the fact that one is moving from $x$ to $x +\epsilon\mu+_{\alpha,\beta}(x)$ as one goes from $U_\alpha$ to  $U_\beta$. 

This construction gives the isomonodromic lift for a connection on $V$; for the same  connection transferred to  $V_0$, there is however some extra choice, in that the connection $\nabla$ thought of as a connection over $V_0$ has poles not only at $D$, but $E$. The monodromy around the latter poles is the identity, and so one could deform infinitesimally by any constant glueing matrix $h_{\alpha,\beta}$  in the $f_\alpha$ trivialisations, as well as moving the poles. This does not change $V$, but can change $V_0$. In particular, the infinitesimal displacement of the pole point of $E$ that one chooses  implies, in essence, a choice of lift of our class $\mu_{\alpha,\beta}$ from $H^1(\Sigma, T_\Sigma(-D))$ to a class $\hat\mu_{\alpha,\beta}$ in $H^1(\Sigma, T_\Sigma(-D-E))$, different from our ``standard" lifts of $\mu$ defined above; set $\nu = \hat\mu -\mu$; it will be a cocycle supported on the punctured disks near $E$.  In the $e_\alpha$ trivialisations, near $E$:
$$ (e_\alpha, x)\leftrightarrow (H_{\alpha,\beta} (1+\epsilon(\hat \mu_{\alpha,\beta}A_{\beta} + h_{\alpha,\beta} )),  e_\beta,x+\epsilon \hat \mu_{\alpha,\beta})$$
Here the infinitesimal changes of gauge $h_{\alpha,\beta}$ satisfy $\nabla h_{\alpha,\beta} = 0$.
What determines the choice of $\nu, h$ is that the deformation one chooses of the connection on $V_0$ should preserve $V_0$, as deformed by the $I_0$ deformation. 
In other words,  one thus must choose  $\nu$  and the constant matrices $g$ at each point of $E$ correctly,  so that  
$$ 0 = <\mu A + \nu A  + h >\in H^1(\Sigma, V_0^*\otimes V_0).$$
 There are exactly enough free parameters to accomplish this: there are $n$ parameters at each point of $E$, one of them given by the displacement of the point, and the $n-1$ others given by the infinitesimal conjugation by $h_{\alpha,\beta}$. This compares well with the $n$ parameters at each point for sections $T$ of ${\ca F}$, and a computation in coordinates shows that these variations are isomorphic, so that they generate the possible variations of the sections $T$ of ${\ca F}$, and so they generate  space $H^1(\Sigma, V_0^*\otimes V_0)$, turning things around and thinking of $V_0$ as being obtained from $V$ by a variable section $T$ of ${\ca F}$. 

Consider now the splitting $I_0$. The deformation preserves $\nabla_0$-constant bases:
 $$ (e_\alpha, x)\leftrightarrow  (H_{\alpha,\beta}e_\beta,x+\epsilon \hat\mu_{\alpha,\beta}),$$
and for the connection $\nabla$, though of as the form $\nabla-\nabla_0$, one wants the deformation to preserve not so much the monodromy, as the ``polar parts" of $V$ over $E$ and of $\nabla$ over $D+E$ in the $\nabla_0$ trivialisation. We must then adjust the connection  matrices $A_\alpha$ for $\nabla$ in the $e_\alpha$-trivialisations  to $A_\alpha(x) +\epsilon a_\alpha(x)$ with $a_\alpha(x)$ representing sections of $ V_0^*\otimes V\otimes K_\Sigma$, satisfying 
$${\ca L}_{\mu_{\alpha,\beta}}(A_\beta(x)) = H_{\alpha,\beta}^{-1}a_\alpha(x)H_{\alpha,\beta} -a_\beta(x)$$
on the overlaps. This is possible: we want ${\ca L}_{\mu_{\alpha,\beta}}(A_\beta(x))= \nabla({\mu_{\alpha,\beta}}(A_\beta(x)))=\nabla_0({\mu_{\alpha,\beta}} $ to be zero in  $H^1(\Sigma, V_0^*\otimes V \otimes K_\Sigma)= H^0(\Sigma, V^*\otimes V_0)^*= \bbc^n$, which it is, since the cocycle $\nabla_0({\mu_{\alpha,\beta}}(A_\beta(x)))$ pairs to zero with any section $\phi$ of $H^0(\Sigma, V^*\otimes V_0 ) = H^0(\Sigma, V_0^*\otimes V_0 )$, as $\phi$ satisfies $\nabla_0(\phi) = 0$.

Recapitulating, the isomonodromic splitting gives infinitesimal variations $\mu_{\alpha,\beta}A_{\beta} $ at $D$ in the transition functions defining $V$, as well as variations $\mu_{\alpha,\beta}A_{\beta} $ at $D$ and $\nu_{\alpha,\beta}A_{\beta} + h_{\alpha,\beta}$ at $E$ in the transition functions defining $V_0$. The variations of the connection over each open set $U_\alpha$are trivial. For the splitting induced by $I_0$, we have trivial variations of the transition functions, and variations of the connection $a_\alpha(x)$  satisfying
${\ca L}_{\mu_{\alpha,\beta}}(A_\beta(x)) = H_{\beta,\alpha}a_\alpha(x)H_{\alpha,\beta} -a_\beta(x) $ on overlaps. Noting again that ${\ca L}_{\mu_{\alpha,\beta}}(A_\beta(x)) = \nabla (\mu_{\alpha,\beta}A_\beta(x))= \nabla_0 (\mu_{\alpha,\beta}A_\beta(x))$, the difference between $I$ and $I_0$ is given by a variation  $\mu_{\alpha,\beta}A_{\beta}$ in the transition functions near $D$ , and variations in the connection $a_\alpha$, satisfying 
\begin{equation}\nabla (\mu_{\alpha,\beta}A_\beta(x) ) = H_{\beta,\alpha}a_\alpha(x)H_{\alpha,\beta} - a_\beta(x)\label{cocycle}\end{equation}
showing that the difference of $I$ and $I_0$ does indeed induce a cotangent vector to ${\ca R}_{k,D,pred}$, given by a cocycle in the hypercohomology (``$\bbh^1$-cocycle") of 
$$V^*\otimes V(-D) \buildrel {\nabla}\over {\longrightarrow} V^*\otimes V\otimes K_\Sigma(D_{red}). $$
remembering that $V_0^*\subset V^*$. We check that the $\bbh^1$-cocycle  \ref{cocycle} is the Hamiltonian vector field $X_{\mu\cdot Q}$. Indeed, let us write the open set  $U_1$ as a disjoint union $U_{1,D}$ and $U_{1,E}$. We restrict to trivialisations over $U_{1,E}$ which are covariant constant for $\nabla$, and to trivialisations over $U_{0,i}$ which are covariant constant for $\nabla_0$. One also, as above, restricts to cocycles in $V^*\otimes V$ representing deformations of the bundle in $H^1(\Sigma,V^*\otimes V)$ supported on $U_{1,E}\cap U_0$ of the form $s_{10} = tT^{-1}$ in the $U_0$ trivialisation, where $t, T$ are holomorphic on the disks of $U_1$ surrounding $E$, and $T$ defines the bundle $V$ as a subbundle of $V_0$. The  tangent vectors $X$ representing deformations of $(V,\nabla)$ of this type are then represented by $\bbh^1$-cocycles
$\nabla(s_{10}) = -b_0$, with $b_0$ a section of $V^*\otimes V\otimes K_\Sigma (D)$ over $U_0\cup U_{1D}$. Under the definition of the symplectic form, one is looking for an $\bbh^1$-cocycle  $\nabla(\phi_{10}) = T\psi_1 T^{-1}-\psi_0, \phi_{10}\in H^0(U_1\cap U_0, V^*\otimes V (-D)), \psi_i\in H^0(U_i, V^*\otimes V \otimes K_\Sigma)$, such that 
\begin{align}dH_{\mu\cdot Q}(X) \equiv \sum_{D}res( 2\mu_{10} tr((\nabla-\nabla_0)b_0) )& =\\ 
 \sum_Dres\ tr (\phi_{10}(2b_0)) +  \sum_E res\ tr  (\phi_{10}(b_0)- s_{10}&(T\psi_1 T^{-1}+\psi_0))\equiv \Omega(X_{\mu\cdot Q},X).\notag \end{align}
Using $T\psi_1 T^{-1} - \nabla(\phi_{10}) =\psi_0$, $\nabla(s_{10}) = -b_0$, and integrating by parts, this becomes
\begin{align}\sum_{D}res( 2\mu_{10} tr((\nabla-\nabla_0)b_0) ))& =\\ 
 \sum_Dres\ tr (\phi_{10}(2b_0)) +& \sum_E res\ tr (2\phi_{10}b_0 - s_{10}(2T\psi_1 T^{-1})).\notag \end{align}

One then sees that setting $\phi_{10} = \mu_{10} (\nabla-\nabla_0)$ near $D$, $\phi_{10} = 0$ near $E$, $\psi_i = a_i$ as above gives the desired result, as $s_{10}(2T\psi_1 T^{-1})= ta_1T^{-1}$ pairs to zero, since $a_1$ lies in $V_0^*\otimes V \otimes K_\Sigma$ and so is of the form $c_1T$, with $c_1$ holomorphic on $U_1$, as a section of $V^*\otimes V \otimes K_\Sigma$. In summary, the difference between the splittings $I$ and $I_0$ is indeed the Hamiltonian vector field.

We can also compare the splittings $I$ and $I_0$ for the deformations of the higher order polar parts over a point in $({\gb}_L\cap{\gt}_L)^*/W_L$. We deal with one pole at a time, let us suppose of order $l$, at $p$. Let $z$ be a coordinate with $z=0$ corresponding to the pole, compatible with the jet $c$ of a coordinate system at our point in ${\ca M}_{g,L}$ . One then has, in a $\nabla_0$-constant basis, the expansion for the matrix $A$ of $\nabla = \nabla-\nabla_0$:
$$A(z) = (A_{-l}z^{-l}+ A_{-l+1}z^{-l+1} +...)dz $$
 The eigenvalues $\lambda_\mu(z)$ of $A(\nabla) = \nabla-\nabla_0$ have expansions $\lambda_\mu(z) = \lambda_{\mu, -l}z^{-l} + \lambda_{\mu, -l+1}z^{-l+1}+...$ and it is the terms of order $1,2,..,l-1$ in these expansions that, in essence, become our Hamiltonians. More invariantly, assuming, as we have, that the leading term $A_{-k}$ is regular and diagonal, we can diagonalise the form $A=\nabla- \nabla_0$:
$$A(z) = Z(z)B(z)Z(z)^{-1}$$
with $Z(z)$ invertible and holomorphic, and $B(z)$ diagonal:
$$B(z) = (B_{-k}z^{-k}+ B_{-k+1}z^{-k+1} +...)dz$$
let  $ \beta $ be a truncated Laurent series with terms of order $-1,-2,..,-l+1$ and values in $\gt$ on a punctured neighbourhood of $p$. Then $d\beta\in ({\gb}_D\cap{\gt}_D)^*$ represents a tangent vector to $({\gb}_L\cap{\gt}_L)^*/W_L$. Let the Hamiltonian function $H_\beta$ be defined by $H_\beta= tr\ res_p (\beta\cdot B)$. Let $X_{\beta\cdot B}$ be the corresponding Hamiltonian vector field. 

We have, around the pole, Stokes sectors  $U_{1i}$; let $U_1$ be the union of these sectors and of the origin, and, away from the pole, let $U_0= \Sigma -\{z>\delta\} $. The terms of order $-l, ...,-2$ of $B$ are fixed over a point in $ ({\gb}_L\cap{\gt}_L)^*/W_L$, and it is these terms that we are deforming. One has over the Stokes sectors $U_{1i}$ near the pole, matrix valued functions $Y_i(z)$, asymptotic to $Z$,  with 
$$A(z) = dY_i(z)Y_i^{-1}(z) + Y_i(z) B(z)_p Y_i^{-1} (z)$$
with $B(z)_p = (B_{-1}z^{-1} + B_{<-1}(z))dz$ the polar part of $B$, representing the class of the connection in $({\gb}_D\cap{\gt}_D)^*/W$, or rather the part of it located at the pole.
To first order ($\epsilon^2=0$), one wants to modify the polar part $B_p$ to $B_p +\epsilon d\beta$, seeing how this lifts to vector bundles $V$ plus connections $\nabla$ under $I$ and $I_0$.

For the isomonodromic splitting $I$, one should again think of our bundle $V$ being given a fixed $\nabla_0$-constant trivialisations on $U_0$ near the pole such that the connection is given by $A(z)$. Now let $Q(z) = Q_{-k+1}z^{-k+1} +...+Q_{-1}z^{-1}$ be such that $ dQ= B_p$. On the $U_{1i}$, one has three trivialisations: 
\begin{itemize}
\item The 0-trivialisation, in which the connection matrix for $\nabla$ vanishes; these trivialisations are related to each other over the different sectors by the Stokes matrices; 
\item   The $B_p$ trivialisation, in which the connection matrix for $\nabla$ is $B_p$; it is related to the 0-trivialisation by the transition function $exp( Q)z^{ B_{-1}}$; 
\item  The $A$-trivialisation, in which the connection matrix is given by $A$, is related to the  $B_p$ trivialisation by the transition matrix $Y_i$. 
The $A$ trivialisation is valid over all of $U_1$, and the 0-trivialisation over $V_i$ is related to the $A$ trivialisation over $U_0$ by a transition matrix $Y_i exp( Q)z^{ B_{-1}} $; in particular varying $Q$ by $Q+\epsilon \beta$ varies the bundle, as well as inducing the variation $B_p +\epsilon d\beta$ in the connection over each $V_i$ in the $B_p$ trivialisation.
\end{itemize}
For the isomonodromic deformations,  the variations of the connections over the $U_{1i}$ are trivial in the $0$ trivialisations, are of the form $B_p +\epsilon d\beta$ in the $B_p$ trivialisations, and are of the form $A + \epsilon Y_id\beta Y_i^{-1}= A + \epsilon \nabla (Y_i\beta Y_i^{-1})$ in the $A$-trivialisation. On the intersection of each $V_i$ with $U_0$, one thus has a hypercohomology cocycle $\mu_{i0} = \beta , \nu_1 = d\beta , \nu_0 =0$ in the $B_p$ trivialisations, and so  $\mu_{i0} = y_i + Y_i\beta Y_i^{-1}, \nu_i = \nabla (y_i + Y_i\beta Y_i^{-1}), \nu_0 =0$ in the $A$-trivialisation.
In our original trivialisations, our variation of the connection is given by $A+\epsilon\cdot 0$ in the $U_0$ trivialisation, $A +\epsilon (\nabla(Z\beta Z^{-1}) + (holom))$ in the $U_1$ trivialisation, with the variation in patching function $Z\beta Z^{-1}$ between the two.

For the splitting  $I_0$, one is interested in ``deforming only the polar part". This amounts to modifying the irregular polar part of the connection $\nabla$ in the $\nabla_0$-flat trivialisation (or equivalently $\nabla-\nabla_0$) by  $d(Z\beta Z^{-1})=\nabla_0(Z\beta Z^{-1})=\nabla(Z\beta Z^{-1}) $ by adding to $\nabla $ a suitable section of $V_0^*\otimes V \otimes K_\Sigma (D)$ with that polar part. This is possible, as the exact sequence 
\begin{equation}
0\rightarrow V_0^*\otimes V \otimes K_\Sigma \rightarrow  V_0^*\otimes V \otimes K_\Sigma (D) \rightarrow  V_0^*\otimes V \otimes K_\Sigma (D)|_D \rightarrow 0,\end{equation}
tells us we can find such a section if the coboundary of $\nabla(Z\beta Z^{-1})_{<1}$ is zero in $H^1(\Sigma, V_0^*\otimes V \otimes K_\Sigma)= H^0(\Sigma, V^*\otimes V_0)^*= \bbc^n$, which it is, since the cocycle $\nabla(Z\beta Z^{-1})$ pairs to zero with any section $\phi$ of $H^0(\Sigma, V^*\otimes V_0 ) = H^0(\Sigma, V_0^*\otimes V_0 )$, as $\phi$ satisfies $\nabla_0(\phi) = 0$.

The difference of the two infinitesimal splittings is now given by a variation  $Z\beta Z^{-1}$ in patching function   and  a variation of connection over $U_0$   given by   $a_0$ with values in $V_0^*\otimes V \otimes K_\Sigma$, and  $a_1$ over $U_1$ which is holomorphic, in short a hypercohomology cocycle $(\mu_{10} = Z\beta Z^{-1}, a_0\in H^0(U_0, V_0^*\otimes V \otimes K_\Sigma), a_1$ holom.) with 
$$\nabla(Z\beta Z^{-1}) = a_1-a_0.$$

We pair this under the Poisson structure with variations of the pairs (bundle, connection) defined by varying the transition functions over open disks $U_{1E}$ , around the support of $E$ by cocycles $s_{10}= tT^{-1}$, of the form given above, with $\nabla s_{0} = -b_0$.
This gives:
$$tr\ res_P (Z\beta Z^{-1}b_0) - \sum_E tr\ res (s_{i0}a_0)$$
The latter terms are zero, as above, since the section $a_0$ takes values in $V_0^*\otimes V \otimes K_\Sigma$, while the first term is the differential of $H_\beta$ applied to the variation of the connection $b_0$, showing that our tangent vector $(\mu_{10}, a_0, a_1)$ is indeed the Hamiltonian vector field $X_{\beta\cdot B}$.

\section{ A symplectic version}

To summarise the situation so far, we have ${\ca V}_{g, L, k}/B_L  \rightarrow  {\ca M}_{g,L}\times ({\gb}_L\cap{\gt}_L)^*/W_L$  with  symplectic fibers, and a Hamiltonian flow on the fiber as  the difference between the lifts of tangent vectors to ${\ca M}_{g,L}\times ({\gb}_L\cap{\gt}_L)^*/W_L$ given by $I$, the isomonodromic lift, and  $I_0$, the lift induced by the isomonodromic deformation of $\nabla_0$. Now we want to make the isomonodromic flow fully Hamiltonian on the whole space. It is essentially a reformulation of the above, but highlights the role of quadratic differentials. 

What we will be doing is a slightly more invariant version of the classical trick of turning a time dependent Hamiltonian flow into an autonomous one by adding a variable. Indeed, if in canonical coordinates one has a time dependent Hamiltonian $H(p, q, t)$, one adds a variable $\rho$ dual to $t$, so that one now has a symplectic form $dp\wedge dq + dt\wedge d\rho$. One then defines an augmented Hamiltonian $\hat H$ by $\hat H(p,q,t,\rho) = H(p,q,t) - \rho$. The flows of $\hat H$ then reproduce the time dependent flow.

The analog of this in our situation is to add to our map ${\ca V}_{g, L, k}/B_L  \rightarrow  {\ca M}_{g,L}\times ({\gb}_L\cap{\gt}_L)^*/W_L$ the dual variables to the moduli:  
\begin{equation}
{\ca X}_{g, L, k } = \{(\Sigma, D, c, V,tr \nabla, \omega, h)/B_L\}
\end{equation}
where $\Sigma, D, V, tr, \nabla$ are as above,  $\omega$ is a quadratic differential with polar divisor bounded by  $D$, and $h$ lies $\gb_L\cap\gt_L$. 
 There is a natural map 
\begin{align}
\Pi: {\ca X}_{g, L, k } & \rightarrow T^*({\ca M}_{g,L}\times ({\gb}_L\cap{\gt}_L)^*)\label{projection} 
\\
(\Sigma, D, c , V, tr, \nabla,\omega, h)& \mapsto (\Sigma, D, c,\omega) ,(irr(\nabla), h))
, \notag 
\end{align}
and a commuting diagram 
\begin{equation}
\begin{matrix}
{\ca X}_{g, L, k } & \buildrel{\Pi}\over{\rightarrow} &T^*({\ca M}_{g,L}\times ({\gb}_L\cap{\gt}_L)^*)\\
P\downarrow&&\downarrow\\
{\ca V}_{g, L, k }/B_L & \rightarrow &{\ca M}_{g,L}\times ({\gb}_L\cap{\gt}_L)^*\\
\end{matrix}\label{diagram}
\end{equation}

The fibers of \ref{projection} are our spaces ${\ca R}^\gamma_{k,D, pred}$. Choosing one of these, and using the symplectic connection $I_0$ decomposes the space locally into a product of symplectic manifolds:

\begin{equation}
{\ca X}_{g, L, k } = {\ca R}^\gamma_{k,D, pred}\times T^*({\ca M}_{g,L}\times ({\gb}_L\cap{\gt}_L))
\end{equation}
and so one has:
\begin{proposition}
${\ca X}_{g, L, k }$ has a symplectic structure determined by $I_0$, for which the map $\Pi$ is Poisson.
\end{proposition}

There is a natural section $S$ of the left hand downward map of \ref{diagram}.  We saw that the Hamiltonians $H_{\mu\cdot Q}$ defined above, depend linearly on $\mu$ at each point $p$ of ${\ca X}_{g, L, k,diag}$, and so define an element $\hat Q=\hat Q(p)$ in the dual $H^0(\Sigma, K^2(D))$ to $H^1(\Sigma, T_\Sigma(-D))$. Similarly, the hamiltonians $H_{\beta\cdot B}$ can be combined into one element $\hat B \in ({\gb}_L\cap{\gt}_L)$, obtained from the restriction of the $B$ defined above to $D$ so that it is  given at the point of multiplicity $l_i$ by the terms of order $1,..., l_i-1$. We then set  

\begin{align}
S: {\ca V}_{g, L, k }/B_L & \rightarrow {\ca X}_{g, L, k},
\\
(\Sigma, D, c , V, tr, \nabla)& \mapsto (\Sigma, D, c , V, tr, \nabla,\hat Q,\hat  B )
.  \notag
\end{align}
Now consider a vector field $Y$ on ${\ca M}_{g,L}\times ({\gb}_L\cap{\gt}_L)$. This of course gives a flow, but it also defines a function $F_Y$ on $T^*({\ca M}_{g,L}\times ({\gb}_L\cap{\gt}_L))$, linear on the fibers, such that the  Hamiltonian flow of $-F_Y$ projects to the flow given by $Y$.
We can now define a function $H_Y$ on ${\ca X}_{g, L, k}$ by 
\begin{equation}
H_Y = F_Y\circ \Pi\circ (S\circ P- Id)
\end{equation}
One has the straightforward:
\begin{proposition}
The Hamiltonian flow of $H_Y$, projected to ${\ca V}_{g, L, k}$, is the isomonodromic lift of the flow of $Y$.
\end{proposition}
This is in essence a reformulation of the result of the previous section. Note that the splitting defined by $I_0$ again intervenes in an important way.


\end{document}